\documentclass{aa}  
\usepackage{natbib}
\bibpunct{(}{)}{;}{a}{}{,} 
\usepackage{graphicx}
\usepackage{txfonts}
\usepackage{xcolor}
\usepackage{amssymb}
\usepackage{amsmath}
\usepackage{siunitx}
\usepackage{hyperref} %
\usepackage{tikz}
\definecolor{lime}{HTML}{A6CE39}
\DeclareRobustCommand{\orcidicon}{%
	\begin{tikzpicture}
	\draw[lime, fill=lime] (0,0) 
	circle [radius=0.16] 
	node[white] {{\fontfamily{qag}\selectfont \tiny ID}};
	\draw[white, fill=white] (-0.0625,0.095) 
	circle [radius=0.007];
	\end{tikzpicture}
	\hspace{-2mm}
}

\foreach \x in {A, ..., Z}{%
	\expandafter\xdef\csname orcid\x\endcsname{\noexpand\href{https://orcid.org/\csname orcidauthor\x\endcsname}{\noexpand\orcidicon}}
}

\usepackage{relsize}
\def\rhh{r_{\textsc{\relsize{-10}{HH}}}}
\DeclareSIUnit\angstrom{\text {Å}}
\begin{document}

   \title{Rovibrational (de-)excitation of H$_{2}$ by He revisited}

   \author{Hubert Jóźwiak\inst{1}\orcidA{} \and
           Franck Thibault\inst{2}\orcidB{} \and
           Alexandra Viel\inst{2}\orcidC{} \and
           Piotr Wcisło\inst{1}\orcidD{} \and
           François Lique\inst{2}\orcidE{}
          }

   \institute{Institute of Physics, Faculty of Physics, Astronomy and Informatics, Nicolaus Copernicus University in Toruń, Grudziądzka 5, 87-100 Toruń, Poland\\
              \email{hubert.jozwiak@doktorant.umk.pl}
         \and
             Univ Rennes, CNRS, IPR (Institut de Physique de Rennes) - UMR 6251, F-35000 Rennes, France
             }

   \date{}

 
  \abstract
   {Collisional (de-)excitation of H$_{2}$ by {helium} plays an important role in the thermal balance and chemistry of various astrophysical environments, making accurate rate coefficients essential for the interpretation of observations of the interstellar medium.}
   {Our goal is to utilize a state-of-the-art potential energy surface (PES) to provide comprehensive state-to-state rate coefficients for {He-induced} transitions among rovibrational levels of H$_2$.}
    {We perform quantum scattering calculations {for} the H$_{2}$-He system and provide state-to-state rate coefficients for {1~089} transitions between rovibrational levels of H$_{2}$ with internal energies up to {$\simeq$15~000~cm$^{-1}$} for temperatures ranging from 20 to 8~000~K.}
   {Our results show good agreement with previous calculations for pure rotational transitions between low-lying rotational levels, but we find significant discrepancies for rovibrational processes involving highly-excited rotational and vibrational states. {We attribute these differences to two key factors: the broader range of intramolecular distances covered by \textit{ab initio} points, and the superior accuracy of the PES, resulting from the utilization of the state-of-the-art quantum chemistry methods, compared to the previous lower-level calculations.}}
   {Radiative transfer calculations performed with {the} new collisional data indicate that the population of rotational levels in excited vibrational states experiences significant modifications, highlighting the critical need for this updated dataset in models of high-temperature astrophysical environments.}

   \keywords{molecular data -- scattering -- rate coefficients}

   \maketitle
%

\section{Introduction}
H$_{2}$ is the most abundant molecule in the Universe, and collisions involving H$_{2}$ play a crucial role in the chemistry and thermal balance of various astrophysical environments. For instance, when a shock wave passes through a molecular gas, heating the medium to temperatures of a few hundred {k}elvins or more, inelastic collisions with H$_2$ determine the rate of cooling of the gas at temperatures in the range of 100 -- 1~000~K~\citep{Aannestad_1973, Shull_1982,Le_Bourlot_1999}. Collisional excitation of H$_2$ and subsequent infrared emission was also one of the important cooling mechanisms in the post-recombination era, after the first molecules were formed. Despite its low abundance in the early Universe, collisions of H$_2$ and its deuterated isotopologue, HD, dominated the cooling process, leading to the gravitational collapse of inhomogeneities in the primordial gas and to the formation of the first stars~\citep{Flower_2001}.

Although less abundant than molecular hydrogen, He is one of the most important collisional partners for H$_2$ (along with H, H$^+$, and H$_2$ itself) {in a large variety of astrophysical environments, such as} diffuse clouds, {dense molecular clouds, protoplanetary disks and in the early Universe}. It is essential to note that typical densities in these environments are not sufficient to achieve Local Thermodynamic Equilibrium (LTE). In such cases, the evolution of molecular populations is determined by two processes: radiative transitions and inelastic collisions. Thus, accurate rate coefficients for collisional processes involving H$_{2}$ are crucial for a reliable interpretation of observations of the interstellar medium. For instance, rate coefficients for rovibrational excitation and deexcitation {of} H$_2$ in collisions with He have been used in calculations of the {cooling rate of the primordial gas}~\citep{Flower_2000, Glover_2008}, analysis of populations of H$_{2}$ in turbulent diffuse interstellar clouds~\citep{Cecchi_Pestellini_2005} and evolution of molecular gas in dark, dense clouds following the passage of C-type shocks~\citep{Yuan_2010, Nesterenok_2018}, as well as in studies of the protoplanetary disks chemistry~\citep{Ruaud_2021}. 

The astrophysical community typically uses the H$_{2}$-He rate coefficients calculated by~\citet{Flower_1998} {(hereafter F98)}, which are available in the BASECOL database~\citep{Dubernet_2013}. These rate coefficients were obtained from quantum scattering calculations performed on the potential energy surface (PES) reported by~\citet{Muchnick_1994}. The authors included rovibrational levels up to ${v=3, j=8}$ in their calculations and provided rate coefficients for temperatures ranging from 100 to 6~000~K. Theoretical rate coefficients for vibrational relaxation from rotational levels in the $v=1$ manifold are in satisfactory agreement (within a factor of 2) with experimental data of~\citet{Dove_1974} and \citet{Audibert_1976}. Independent quantum scattering calculations performed on the same PES by~\citet{Balakrishnan_1999a, Balakrishnan_1999b}, but using a different approach to performing a rovibrational average of the PES led to a good agreement with the results of~\citet{Flower_1998} for transitions involving $|\Delta j|=0, 2$, and 4. Importantly, the calculations of \citet{Balakrishnan_1999b} extended the original dataset of \citet{Flower_1998} to $v=4, 5$ and $6$.

The quality of theoretical rate coefficients is largely determined by the potential energy surface (PES) employed. Since the works of~\citet{Flower_1998} and \citet{Balakrishnan_1999a,Balakrishnan_1999b}, new PESs have been published that should, in principle, provide a more accurate description of H$_{2}$-He interaction energies. \citet{Lee_2005} conducted a comparative study between the rate coefficients derived from the \citet{Muchnick_1994} (MR) PES and the then most recent PES reported by \citet{Boothroyd_2003} (BMP). Surprisingly, they discovered that the total quenching rate from the $v=1, j=0$ state calculated with the BMP PES was three orders of magnitude larger than that calculated with the MR PES, and the experimental data reported by~\citet{Audibert_1976}. In a joint experimental and theoretical study of low-temperature (between 22 and 180~K), inelastic collisions of H$_{2}$ and He,~\citet{Tejeda_2008} found that the rate coefficients for pure rotational deexcitation, $k_{v=0, j=2\rightarrow v'=0,j'=0}$, obtained from quantum scattering calculations based on the MR PES agreed better with experimental data than the ones derived from the BMP PES. The two papers confirmed the quality of the MR PES and, indirectly, the rate coefficients calculated by~\citet{Flower_1998}.

However, studies comparing rate coefficients may not be sensitive enough to discern the differences between various PESs. A much more stringent test of the PES can be performed by comparing spectroscopic parameters of molecular lines, such as pressure broadening and shift coefficients~\citep{Thibault_2016}. Recently, we have shown that the most recent H$_{2}$-He PES~\citep{Thibault_2017} (hereafter BSP3) leads to a subpercent agreement between experimental and theoretical line profiles of molecular hydrogen perturbed in collisions with He~\citep{Slowinski_2020}. In particular, we have demonstrated that quantum scattering calculations based on the older PESs, such as the improved version of the~\citet{Muchnick_1994} (mMR), fail to reproduce the cavity-enhanced spectra of the 3-0 S(1) and 2-0 Q(1) He-perturbed lines of H$_{2}$~\citep{Slowinski_2022}. The most recent PES has been successfully used in comprehensive calculations of line-shape parameters for He-perturbed H$_{2}$~\citep{Jozwiak_2018} and HD~\citep{Thibault_2020,Stankiewicz_2020} lines, which provided a basis for the construction of a new generation of line-shape parameters databases based on \textit{ab initio} calculations~\citep{Wcislo_2021,Stankiewicz_2021}. The BSP3 PES has also been used in studies of stereodynamics of cold He-HD~\citep{Morita_2020, Morita_2020b} and He-D$_{2}$~\citep{Jambrina_2022} collisions.

In this paper, we use the most recent PES~\citep{Thibault_2017} to perform quantum scattering calculations {for the} H$_{2}$-He system and to revise the state-to-state rate coefficients calculated by~\citet{Flower_1998}. {The paper is organized as follows.} In Sec.~\ref{sec:methods} we briefly describe the quantum scattering calculations. We present examples of cross-sections and rate coefficients and compare the results with available theoretical and experimental data in Sec.~\ref{sec:results}. In Sec.~\ref{sec:discussion} we discuss the potential implications of our results on astrophysical models {before concluding remarks given in Sec.~\ref{sec:conclusions}}. {The complete dataset with all state-to-state coefficients will be available online from the BASECOL~\citep{Dubernet_2013} website.}

\section{Methods}
\label{sec:methods}
We study the inelastic collisions in the H$_{2}$-He systems by solving the close-coupling equations in the body-fixed (BF) frame of reference. The theory of non-reactive scattering in systems with arbitrary angular momenta in the BF frame was developed by~\citet{Launay_1977} and was recently recalled by some of us in the context of scattering calculations for He-perturbed shape of HD rovibrational resonances~\citep{Stankiewicz_2020}. We choose the BF frame of reference because the coupling matrix exhibits a predominantly block-diagonal structure with blocks interconnected by centrifugal terms~(see Fig.~3 in~\citet{Rabitz_1975}), which significantly reduces computational time and memory requirements. 

We employ the H$_{2}$-He potential energy surface reported by~\citet{Thibault_2017}, which extends the PES calculated by~\citet{Bakr2013}. The PES was computed using the coupled-cluster method with single, double, and perturbative triple excitations [CCSD(T)], supplemented by full configuration interaction corrections. In comparison to the previous version, which covered H--H distances in the range of ${\rhh\in [1.1, 1.7]~a_{0}}$, this PES encompasses a broader range of intramolecular H--H distances (${\rhh\in [0.65, 3.75]~a_{0}}$), a critical aspect for accurately studying inelastic transitions between excited vibrational states of H$_{2}$. 

The PES is expanded in the Legendre polynomials
\begin{equation}
\label{eq:potential-expansion}
    V(R, \rhh, \theta) = \sum_{\lambda = {0, 2, 4, 6}} A_{{\lambda}} (R, \rhh) P_{\lambda} (\cos{\theta}),
\end{equation} 
where $R$ denotes the distance between He and the center of mass of H$_{2}$, and $\theta$ is an angle between the intra- and intermolecular axis. {Since H$_{2}$ is a homonuclear molecule, $\lambda$ takes only even values}. The dependence of the expansion coefficients on the stretching coordinate, $\rhh$, is ruled out by averaging the PES over {rovibrational} wave functions of isolated H$_{2}$ molecule, $\chi_{vj}(\rhh)$,
\begin{equation}
\label{eq:potential-rovibaverage}
    v_{{\lambda},v,j, v',j'} (R)= \int \mathrm{d}\rhh \chi_{vj}(\rhh) A_{{\lambda}} (R, \rhh) \chi_{v'j'}(\rhh)  .
\end{equation}
We use a total number of 703 {coupling} terms, {$v_{{\lambda},v,j, v',j'} (R)$}, which describe couplings between all 37 (see below) rovibrational states included in the calculations.

We recall that the close-coupling equations are diagonal with respect to the total angular momentum ($J$) and parity ($p$), i.e. they can be solved for each value of $J$ and $p$ separately. Since the PES does not couple states with different nuclear spins ($I$), we consider scattering of \textit{para}-H$_{2}$ ($I=0$, rotational states with even $j$ values) and \textit{ortho}-H$_{2}$ ($I=1$, rotational states with odd $j$ values) separately. The coupled equations are solved using renormalized Numerov's algorithm implemented in the BIGOS code~\citep{BIGOS}. The log-derivative matrix is transformed to the space-fixed (SF) frame of reference at a sufficiently large value of $R$, and the boundary conditions, imposed on the scattering wave functions allow us to obtain the S-matrix elements. The state-to-state cross-sections are calculated from the S-matrix elements as follows
\begin{align}
    \begin{split}
        \sigma_{vj \rightarrow v'j'}(E_{\rm{kin}}) = & \frac{\pi}{k_{vj}^{2}(2j+1)} \\ &\times \sum_{J, l, l'} (2J+1) \Bigl|\delta_{vv'}\delta_{jj'}\delta_{ll'} -S_{vjl\rightarrow v'j'l'}^{J}(E)\Bigr|^{2},
    \end{split}
\end{align}
where $v$ and $j$ denote pre-collisional vibrational and rotational quantum numbers of H$_{2}$, $l$ is the angular momentum of the relative motion in the colliding system, $E = E_{\rm{kin}} + E_{vj}$ is the total energy (the sum of the relative kinetic energy and the internal energy of H$_{2}$ in the given rovibrational state), ${k_{vj} ={\hbar^{-1}}\sqrt{2\mu(E-E_{vj})}}$, $\mu$ is the reduced mass of the H$_{2}$-He system, {and $\hbar$ is the reduced Planck constant}. Primed symbols denote post-collisional values. The sum over $J$ is truncated at a value $J_{\rm{max}}$ large enough to ensure convergence of the cross-sections at the level of $10^{-4}$~\si{\angstrom}$^{2}$. The range of the sum over $l$ (and $l'$) is determined by the triangular rule resulting from adding {two} angular momenta $\mathbf{j}+\mathbf{l}=\mathbf{J}$ (and $\mathbf{j}'+\mathbf{l}'=\mathbf{J}$).

 We are interested in rovibrational transitions between H$_{2}$ levels with internal energies lower than 15~000~cm$^{-1}$. The cross-sections are calculated for total energies ranging between 10$^{-6}$ and  40~000~cm$^{-1}$ with various energy steps, in order to accurately describe the channel-opening effects and resonances. We tested the convergence of the cross-sections with respect to the parameters of the propagator and the size of the rovibrational basis. Ultimately, the range of the propagation was set to $R_{\rm{min}}=1\,a_{0}$, and $R_{\rm{max}}=100\,a_{0}$ with 15~steps per half-de Broglie wavelength, corrected for the depth of the isotropic part of the PES (approximately~10~cm$^{-1}$). In all scattering calculations, the basis set involved 37 rovibrational levels (up to the $v=2, j=14$  {for} \textit{para}-H$_{2}$ and $v=5, j=1$ {for} \textit{ortho}-H$_{2}$). At the largest total energy considered in this work (40~000~cm$^{-1}$), this size of the basis set ensured convergence better than 10\% {with respect to the fully-converged basis}\footnote{The fully-converged basis included levels up to $v=5, j=6$ and $v=5, j=7$ for \textit{para}-H$_{2}$-He and \textit{ortho}-H$_{2}$-He collisions, respectively.} for 539 (out of 729) in the \textit{para}-H$_{2}$ case and 520 (out of 676) in the \textit{ortho}-H$_{2}$ case cross-sections related to transitions between levels with internal energies lower than 15~000~cm$^{-1}$, {leading to the total number of 1~089 transitions}.

We calculate thermal rate coefficients by averaging the state-to-state cross-sections over the distribution of relative kinetic energy
\begin{equation}
    k_{vj \rightarrow v'j'} (T) = { \langle{v}_{r}\rangle  \Biggl(\frac{1}{k_{\rm{B}}T}\Biggr)^{2}} \int_{0}^{\infty} \sigma_{vj \rightarrow v'j'}(E_{\rm{kin}})E_{\rm{kin}} e^{-E_{\rm{kin}}/k_{\rm{B}}{T}} \mathrm{d}E_{\rm{kin}}, 
\end{equation}
{where $\langle{v}_{r}\rangle  = \sqrt{8k_{\rm{B}}T/(\pi \mu )}$ is the mean relative speed of the colliding partners {at a given temperature}.}  The range of total energies covered by the scattering calculations (from 10$^{-6}$ to 40~000~cm$^{-1}$) allows us to determine thermal rate coefficients ranging from 20 to 8~000~K for {1~089} {considered} transitions between rovibrational levels with internal energy lower than 15~000~cm$^{-1}$.

 \section{Results}
\label{sec:results}
In this Section, we discuss the calculated cross-sections and thermal rate coefficients.

\subsection{Pure rotational ($v'=v$) (de-)excitation}
\begin{figure}[!ht]
    \centering
    \includegraphics[width=\linewidth]{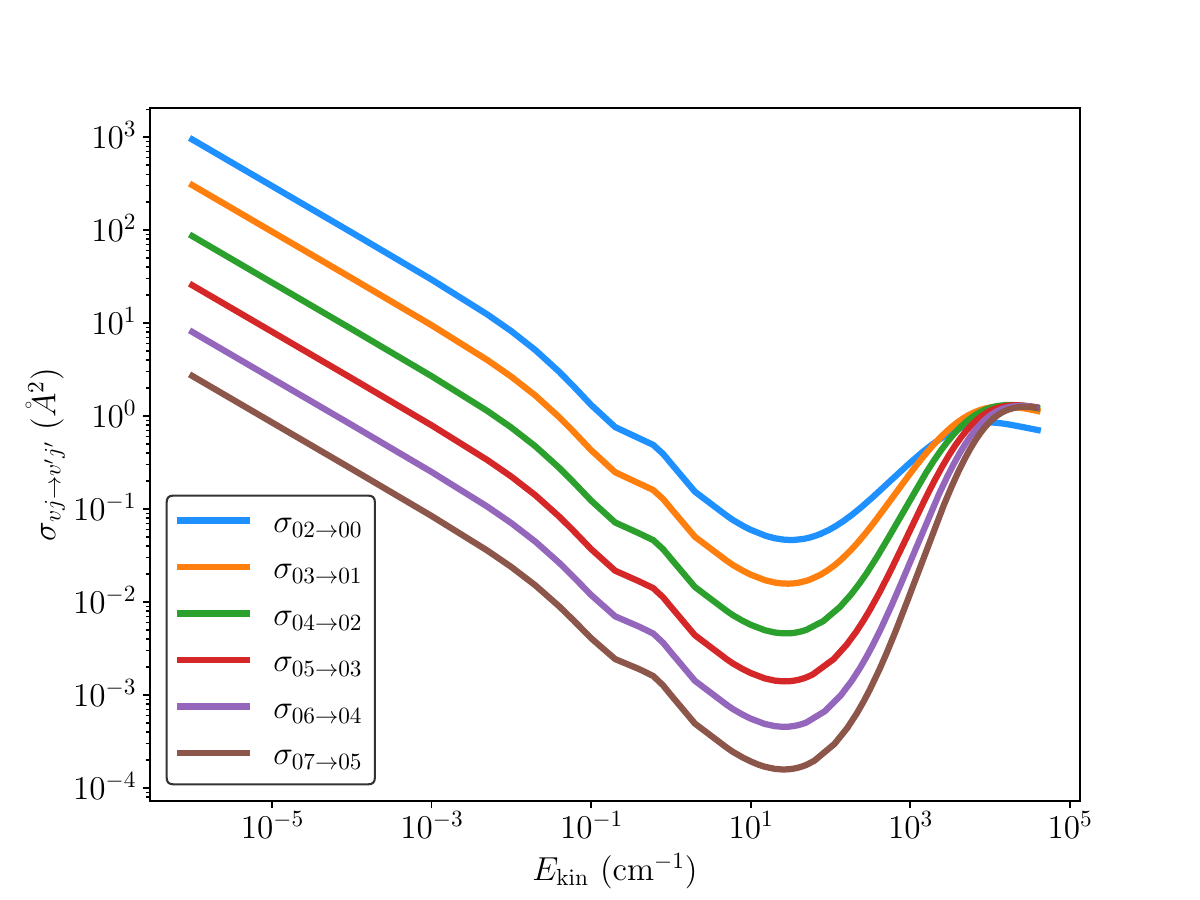}\vspace{-0.1cm}
    \includegraphics[width=\linewidth]{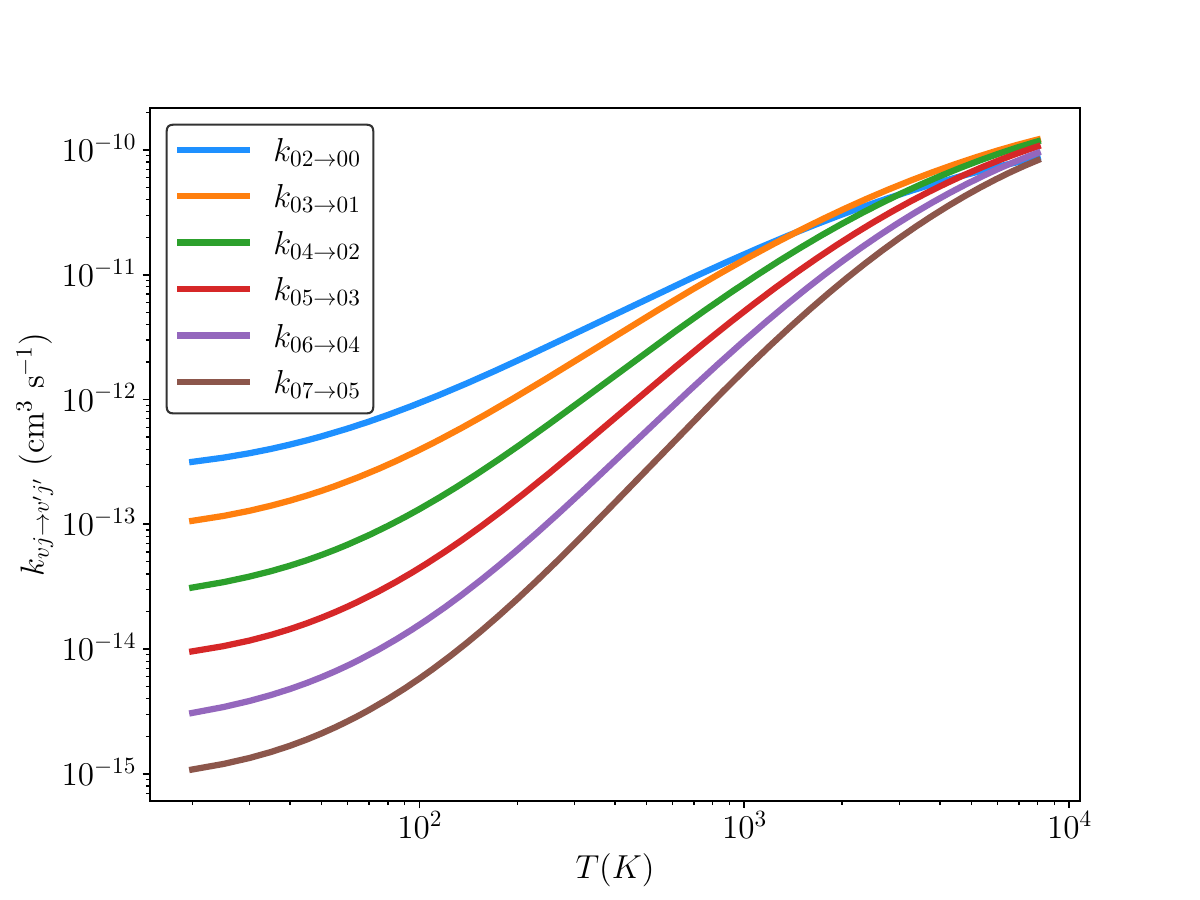}
    \caption{State-to-state cross-sections (top panel) and corresponding thermal rate coefficients (bottom panel) for pure rotational deexcitation in $v=0$ with $\Delta_{j}=-2$.}
    \label{fig1}
\end{figure}
We begin with the inelastic processes between rotational levels within the ground vibrational state, which are the most populated states in H$_{2}$ in the {100 -- 1000~K temperature range}. Since the PES does not couple \textit{ortho}- and \textit{para}-H$_{2}$, $\Delta {j} = \pm 1$ transitions are forbidden. The dominant inelastic processes are the ones that change the rotational quantum number by $\Delta j = \pm 2$. These are {mostly} driven by the $\lambda = 2$ term in the expansion in Eq.~\eqref{eq:potential-expansion}, which is the largest anisotropic term in the expansion of the PES. Fig.~\ref{fig1} presents the state-to-state cross-sections for pure rotational deexcitation with $\Delta j = -2$ of H$_{2}$ upon collisions with helium up to $j = 7$. At ultra-low kinetic energies (below 10$^{-3}$~cm$^{-1}$, the cross-sections obey the Wigner threshold law~(\citet{Wigner_1948}) for deexcitation cross-section ($\sigma \sim k^{-1/2}$). As the kinetic energy increases, more partial waves contribute to the inelastic scattering, the cross-sections pass through a minimum near 20~cm$^{-1}$ and gradually increase at higher kinetic energies. For the majority of considered kinetic energies, the cross-sections decrease with increasing initial rotational state. This property is translated to thermal rate coefficients (bottom panel in Fig.~\ref{fig1}), which also decrease with increasing rotational quantum numbers. This is related to the fact that the rotational spacing between levels increases with $j$, thus {reducing} the deexcitation process.

\begin{figure}[!ht]
    \centering
    \includegraphics[width=\linewidth]{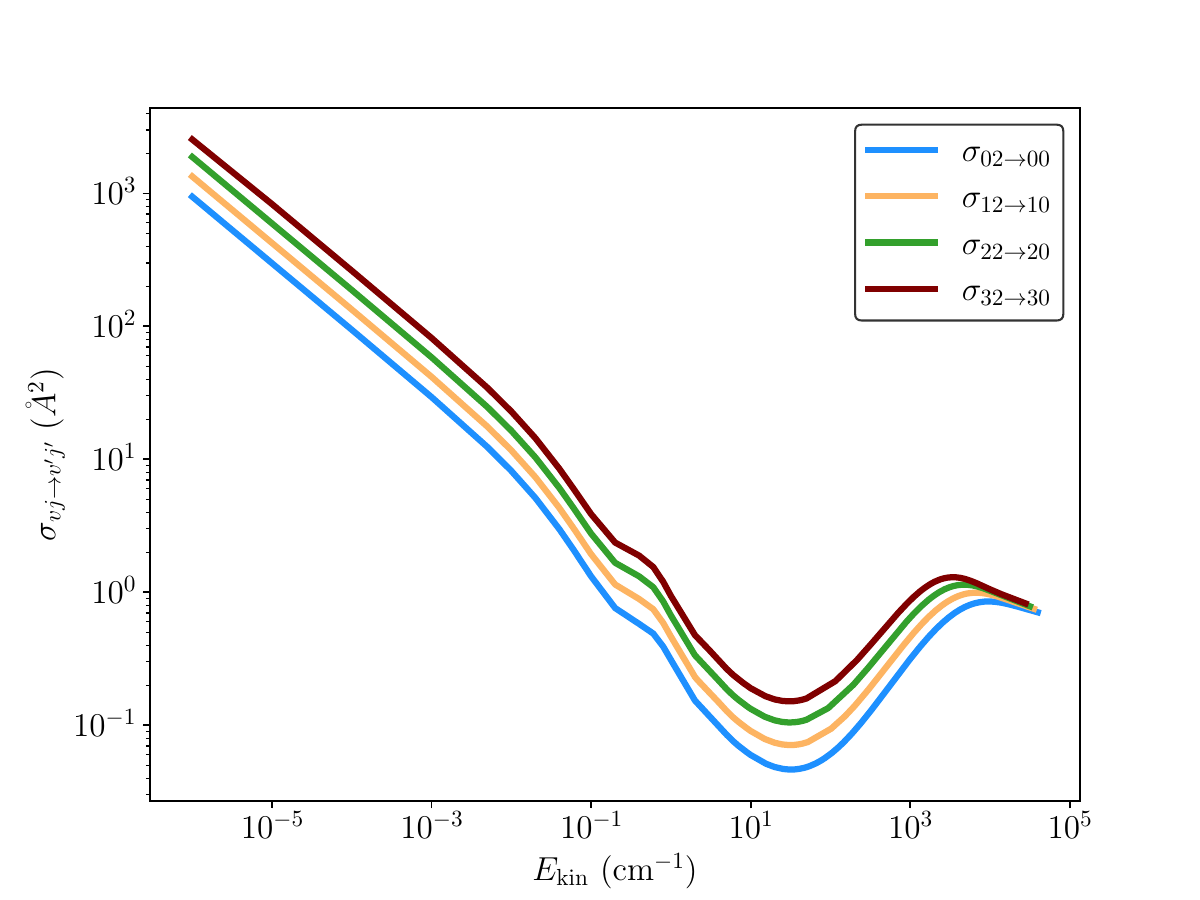}\vspace{-0.1cm}
    \includegraphics[width=\linewidth]{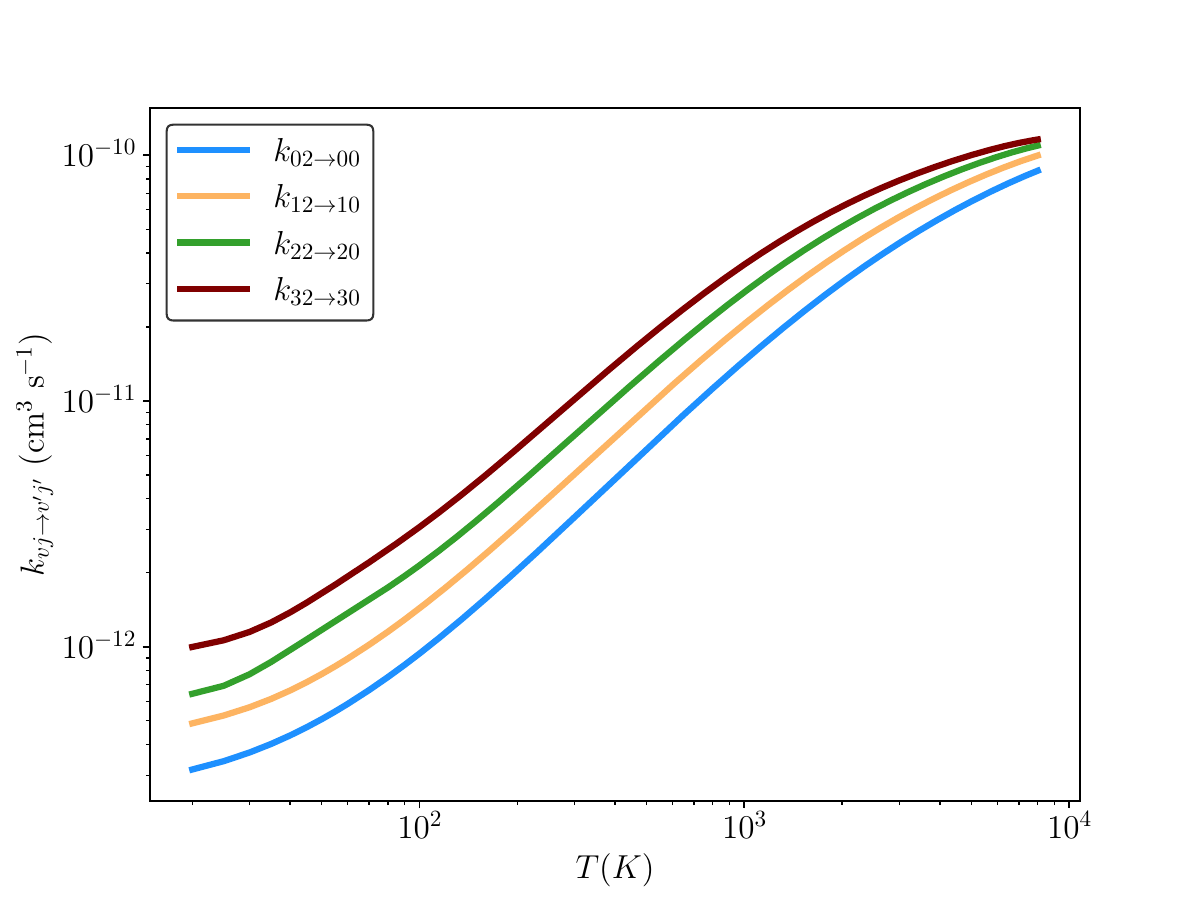}
    \caption{State-to-state cross-sections (top panel) and corresponding thermal rate coefficients (bottom panel) for $v, j=2\rightarrow v, j'=0$ deexcitation {with $v = 0, 1, 2, 3$}.}
    \label{fig2}
\end{figure}
Pure rotational deexcitation (with $\Delta j = -2$) within \textit{excited} vibrational states looks similar to the deexcitation within $v=0$. The only difference is that the cross-sections (and corresponding rate coefficients) increase with the increasing value of $v$. Fig.~\ref{fig2} presents a comparison between the $\sigma_{v, 2 \rightarrow v, 0}$ (in the top panel) and $k_{v, 2 \rightarrow v, 0}$ (in the bottom panel) for different vibrational quantum numbers. The observed trend is explained by the fact that the energy interval between rotational states decreases with increasing $v$, thus enhancing rotational transitions within the same vibrational level. {This is in line with} an increased contribution from inelastic processes to pressure broadening of rovibrational lines {as $v$ increases}~(\citet{Hartmann_2021}), {as} observed in He-perturbed Q~(\citet{Thibault_2017}), and S and O~(\citet{Jozwiak_2018}) lines of H$_{2}$.

\subsection{Rovibrational (de-)excitation}
\begin{figure}[!ht]
    \centering
    \includegraphics[width=\linewidth]{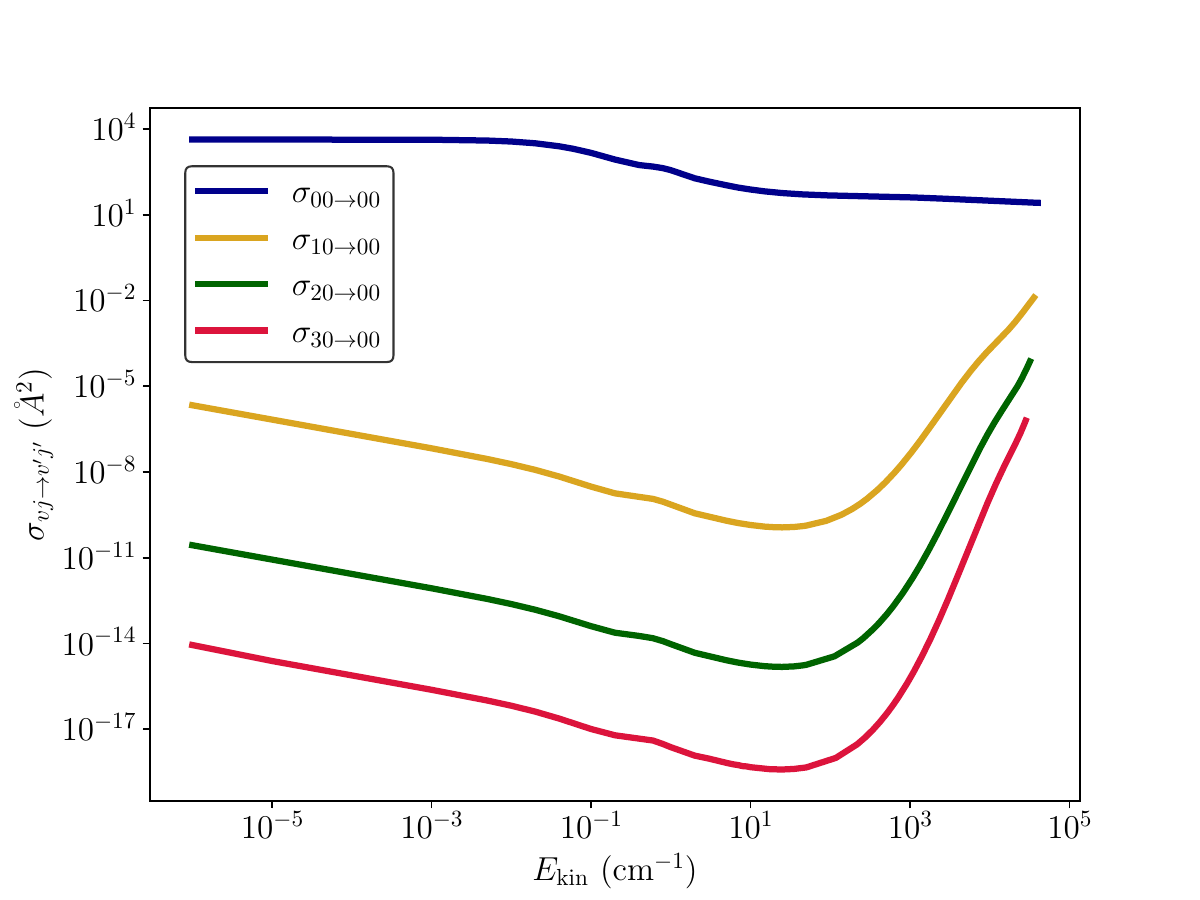}\vspace{-0.1cm}
    \includegraphics[width=\linewidth]{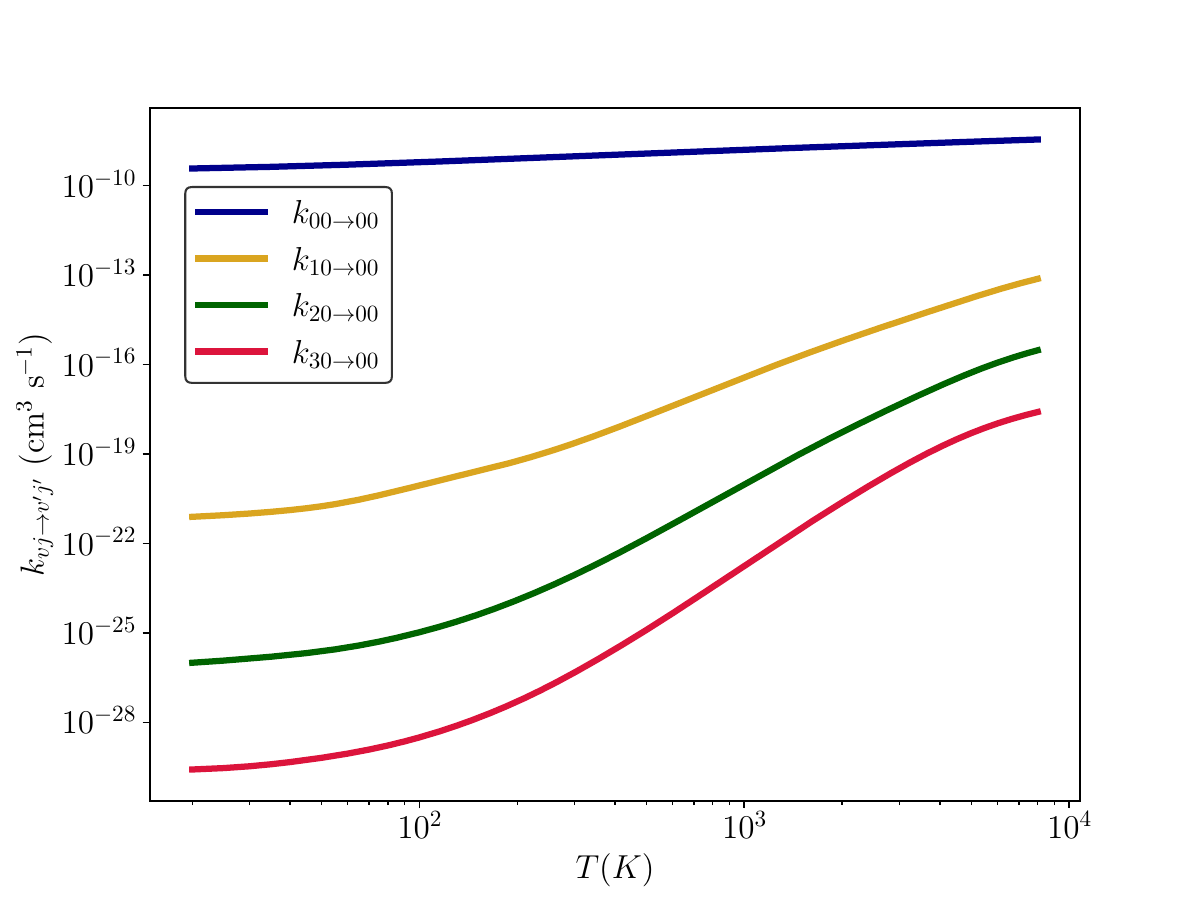}
    \caption{State-to-state cross-sections (top panel) and corresponding thermal rate coefficients (bottom panel) for $v, j=0\rightarrow v'=0, j'=0$ deexcitation {with $v = 0, 1, 2, 3$}.}
    \label{fig3}
\end{figure}
Figure~\ref{fig3} presents a comparison of the state-to-state cross-sections and corresponding rate coefficients for $\Delta v, \Delta j = 0$ transitions. For reference, we put the elastic scattering cross-section in the $v=0, j=0$ state (dark blue color), which is at least three orders of magnitude larger than the cross-sections for vibrationally inelastic processes. We recover the expected asymptotic behavior of the elastic cross-section at ultra-low kinetic energies, {where it}  converges to the constant value. The dependence of $v$-changing cross-sections follows a trend similar to pure rotational deexcitation, with a minimum near 20~cm$^{-1}$, but the increase in the orders of magnitude of the cross-sections at high ($>10^{3}$~cm$^{-1}$) is more drastic. For instance, the $\sigma_{1,0\rightarrow 0,0}$ cross-section increases by 8 orders of magnitude for kinetic energies in the range 20 to 40~000~cm$^{-1}$. Overall, the cross-sections and the corresponding thermal rate coefficient decrease rapidly with increasing $\Delta v$, owing to the fact that the off-diagonal (in terms of vibrational quantum numbers) radial coupling terms of the potential decrease with $\Delta v$. This trend is transferred to thermal rate coefficients. We note that the rate coefficients for vibrational-state-changing collisions are at least 3 orders of magnitude {smaller} (at $T=8000$~K) than the rate coefficients for pure rotational deexcitation presented in the top panels of Fig.~\ref{fig1} and~\ref{fig2}.

\begin{figure}[!ht]
    \centering
    \includegraphics[width=\linewidth]{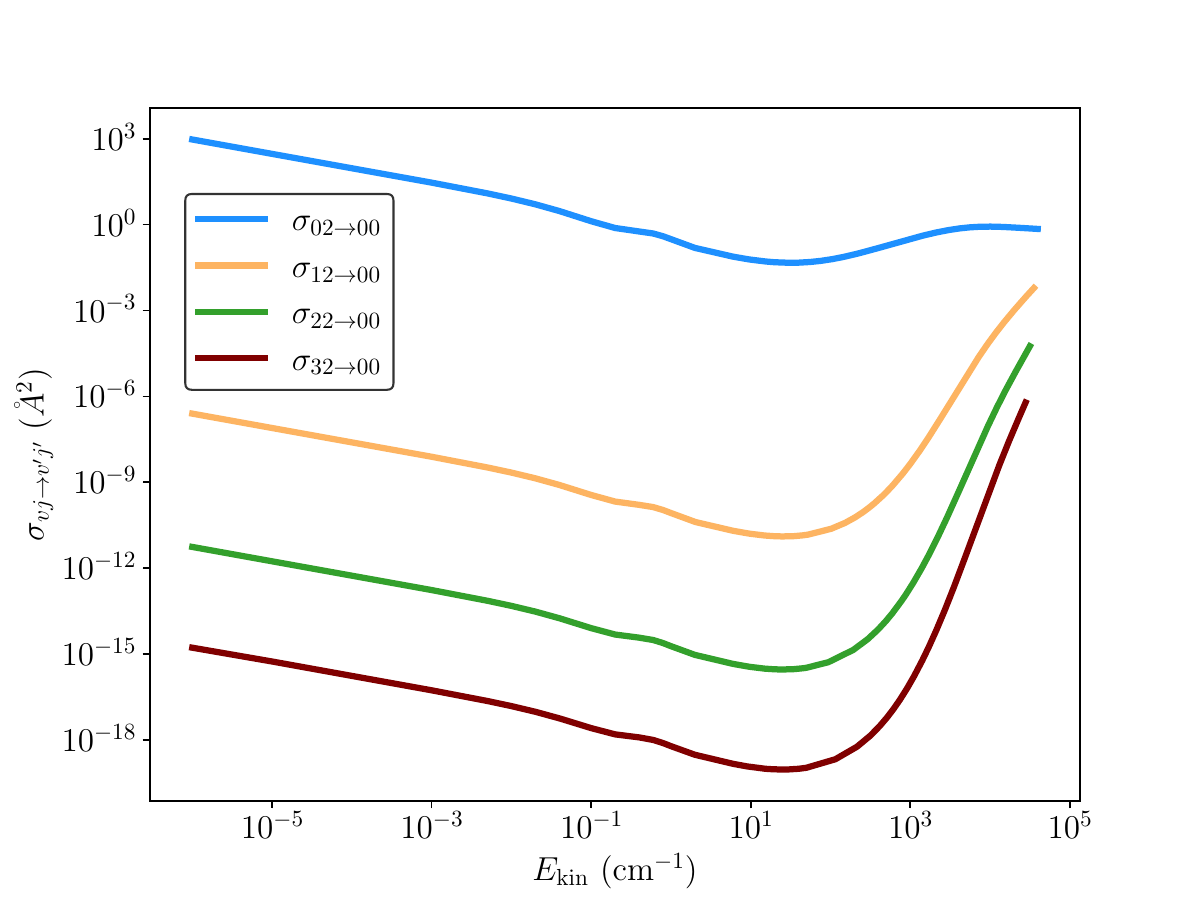}\vspace{-0.1cm}
    \includegraphics[width=\linewidth]{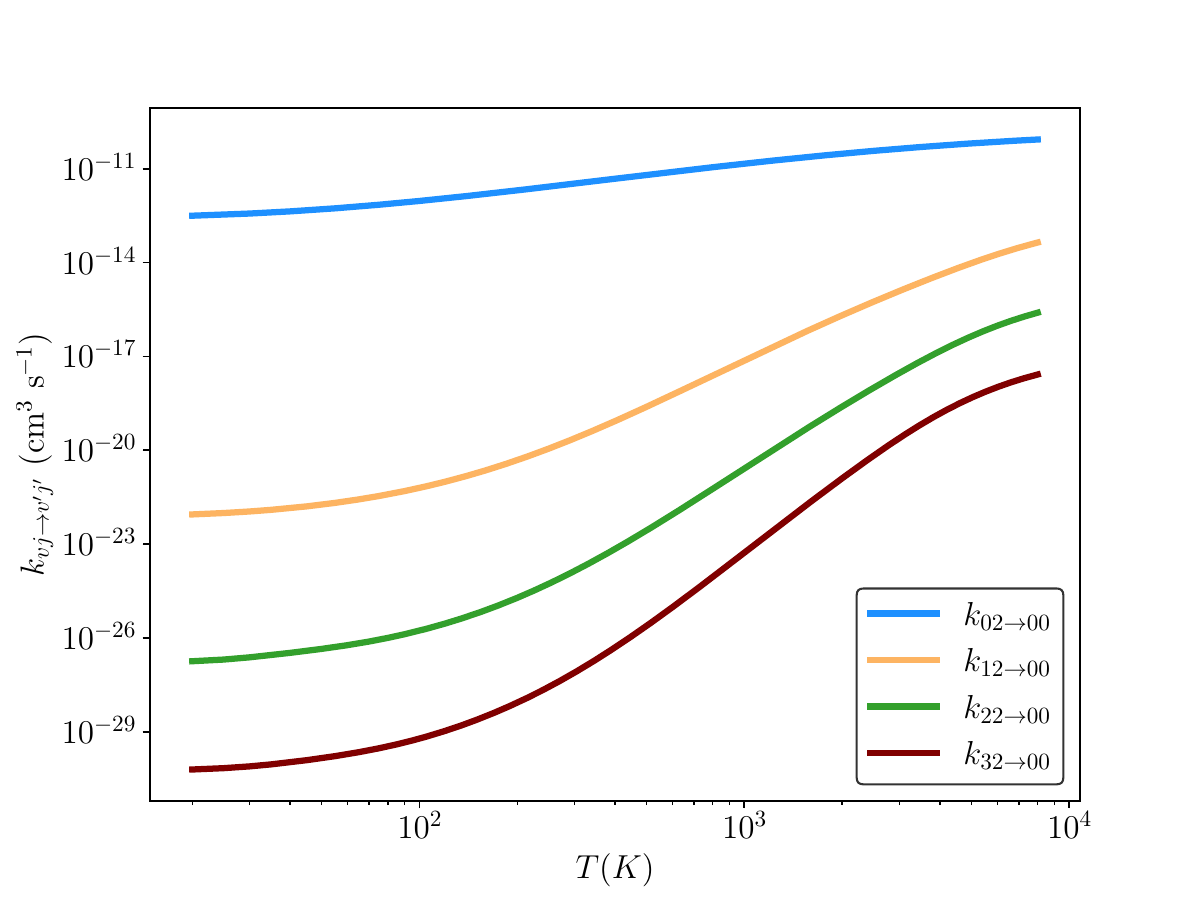}
    \caption{State-to-state cross-sections (top panel) and corresponding thermal rate coefficients (bottom panel) for $v, j=2\rightarrow v'=0, j'=0$ deexcitation from the four considered vibrational states.}
    \label{fig4}
\end{figure}
Rate coefficients for simultaneous quenching of both rotational and vibrational excitation are even smaller. Figure~\ref{fig4} presents state-to-state cross-sections and thermal rate coefficients for the ${v, j=2\rightarrow v'=0, j'=0}$ transitions. Contrary to {${v, j=0\rightarrow v'=0, j'=0}$} transitions from Fig.~\ref{fig3}, driven {only} by $\lambda=0$ terms which are off-diagonal in vibrational quantum numbers, transitions in Fig.~\ref{fig4} owe their strength to the anisotropic $\lambda=2$ terms which are off-diagonal in $v, v'$. Due to the reasons discussed above, the cross-sections and rate coefficients rapidly decrease with increasing $\Delta v$.

For the sake of brevity, we do not discuss analogous (${v, j=1\rightarrow v'=0, j'=1}$ and ${v, j=3\rightarrow v'=0, j'=1}$) transitions in \textit{ortho}-H$_{2}$. The dependence of the cross-sections and rate coefficients on $\Delta v$ and the overall relation with respect to kinetic energy and temperature remains {similar}.

\subsection{Comparison with previous results}
\begin{figure*}[!ht]
    \centering
    \includegraphics[width=0.45\linewidth]{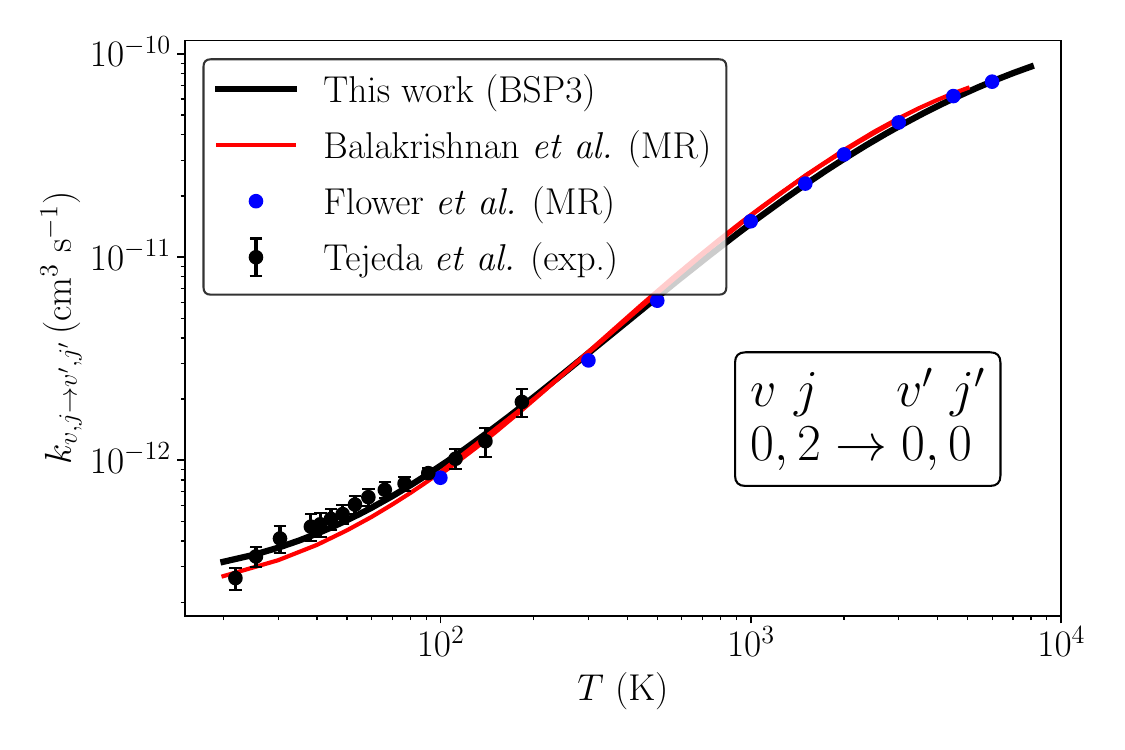}
    \includegraphics[width=0.45\linewidth]{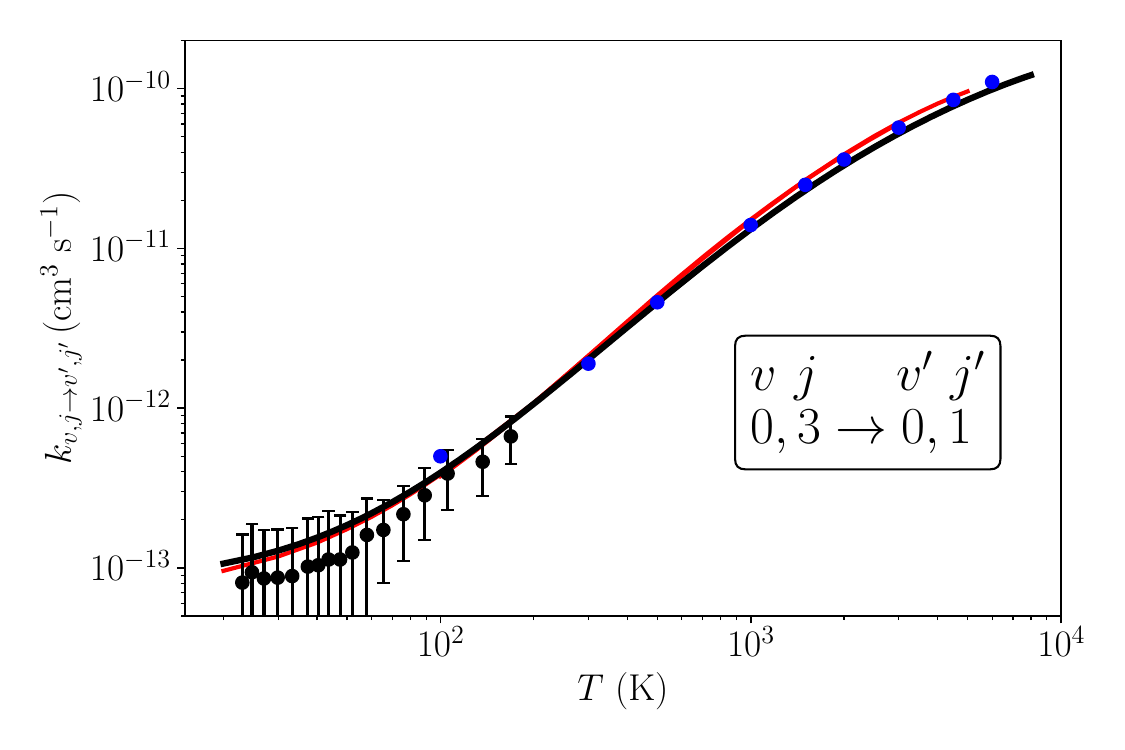}\\\vspace{-0.61cm}
    \includegraphics[width=0.45\linewidth]{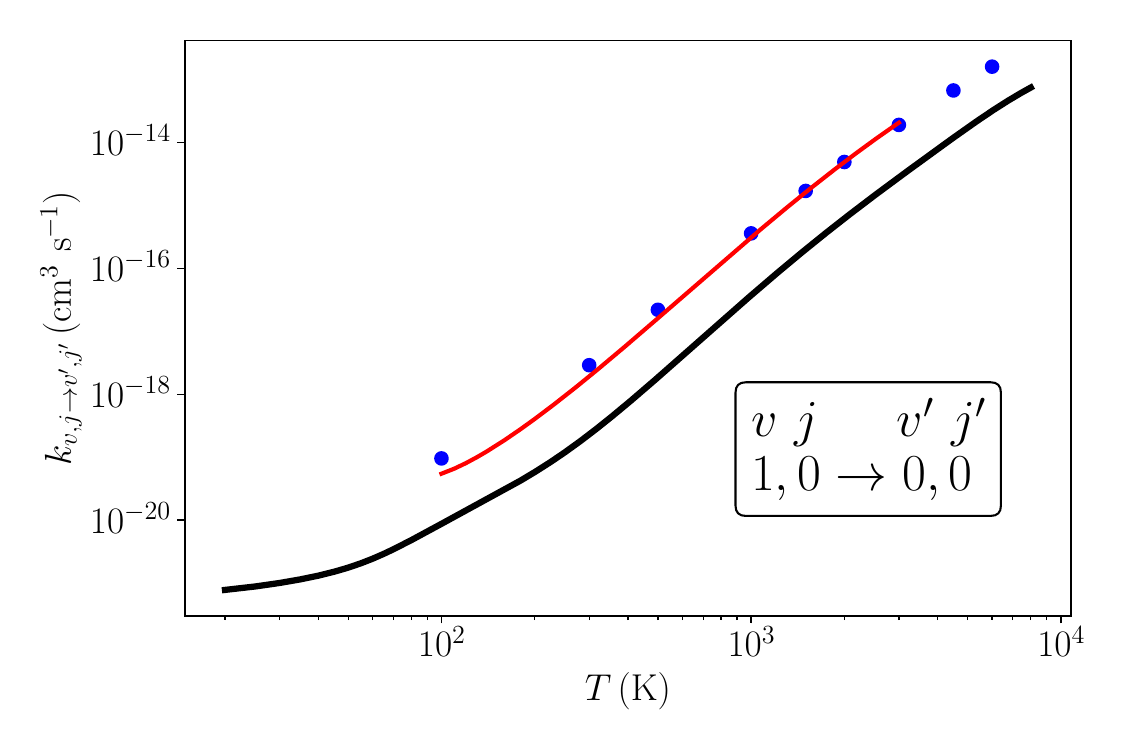}
    \includegraphics[width=0.45\linewidth]{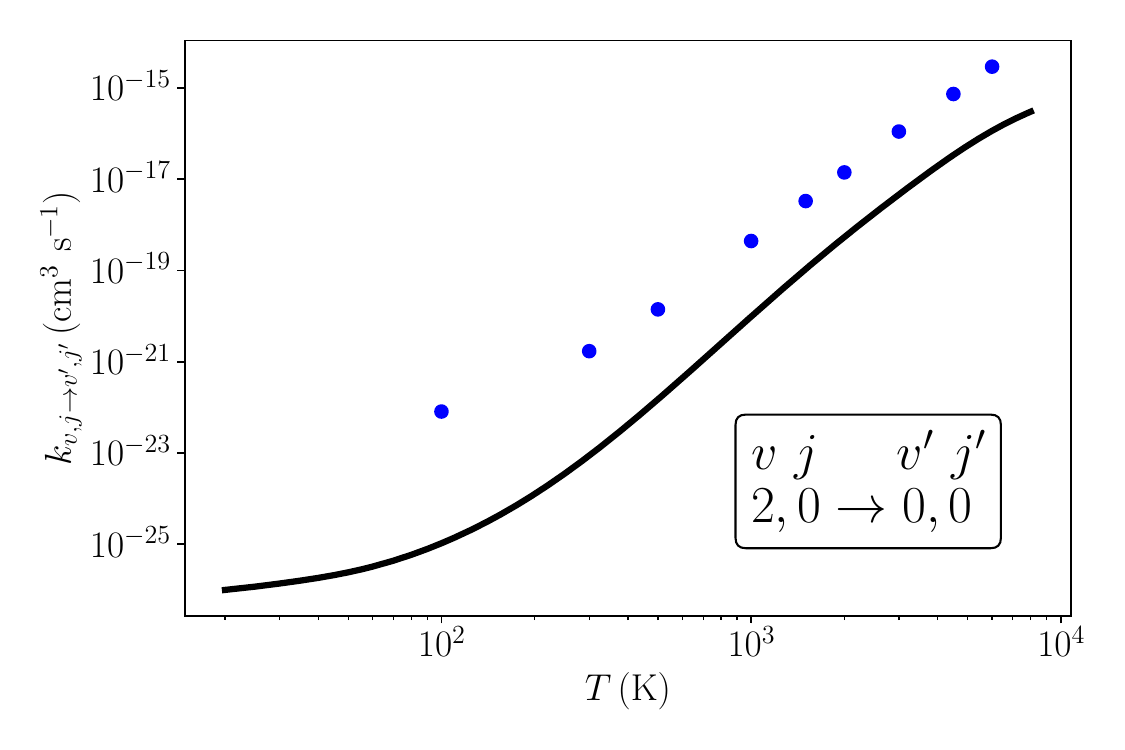}
    \caption{Comparison between the rate coefficients for pure rotational $v=0, j=2 \rightarrow v=0, j=0$ (top left panel), $v=0, j=3 \rightarrow v=0, j=1$ (top right panel), and rovibrational $v=1, j=0 \rightarrow v'=0, j'=0$ (bottom left panel) and $v=2, j=0 \rightarrow v'=0, j'=0$ (bottom right panel) deexcitation provided in this work and available literature data.}
    \label{fig5}
\end{figure*}
We compare the {new} rate coefficients with available literature data. Due to the large size of the dataset we provide, we discuss a few representative rovibrational transitions that highlight the differences and similarities between our results and previous findings. 

We begin with the comparison of rate coefficients for pure rotational deexcitation of H$_{2}$ in the $v=0$ state upon collisions with He. In the top left ($k_{0,2\rightarrow 0,0}$) and right ($k_{0,3\rightarrow 0,1}$) panels of Fig.~\ref{fig5} we gathered theoretical rate coefficients calculated in this work, those reported by~\citet{Flower_1998} and~\citet{Balakrishnan_1999a}, as well as experimental results~\citep{Tejeda_2008} determined using Raman spectroscopy in supersonic jets of H$_{2}$-He mixtures. For theoretical results, we use abbreviations of the names of PESs employed in quantum scattering calculations. We note that both~\citet{Flower_1998} and~\citet{Balakrishnan_1999a} performed quantum scattering calculations on the same PES, but used different approach to the rovibrational average in Eq.~\eqref{eq:potential-rovibaverage}. \citet{Flower_1998} approximated the H$_{2}$ wave functions as harmonic oscillator wave functions (independent of $j$), while~\citet{Balakrishnan_1999a,Balakrishnan_1999b} obtained $\chi_{v j}(\rhh)$ as expansions in Hermite polynomials using the potential energy curve for H$_{2}$ taken from~\citet{Schwenke_1988}.

We observe a very good agreement with all previous theoretical calculations, both in the cold ($\simeq 10$~K) regime, as well as at temperatures of an order of $1000$~K. For the ${v=0,j=2\rightarrow v'=0,j'=0}$ transitions, our low-temperature results reproduce the experimental datapoints slightly better than other theoretical rate coefficients. We note that~\citet{Tejeda_2008} also reported results of extensive scattering calculations for the two transitions using the MR PES, its modified (mMR) version~\citep{Boothroyd_2003}, and the BMP PES. We do not add these results to the top panels in Fig.~\ref{fig5} to maintain the readability of the plot. For the ${v=0,j=2\rightarrow v'=0,j'=0}$ deexcitation, using MR and mMR PESs,~\citet{Tejeda_2008} obtained a perfect agreement with the results of~\citet{Balakrishnan_1999a}, but rate coefficients obtained using the BMP PES deviated significantly from both theoretical and experimental datapoints. A similar agreement between theoretical calculations from the MR and mMR PESs was found for rate coefficients for the ${v=0,j=3\rightarrow v'=0,j'=1}$ deexcitations, with results derived from the BMP PES being $\simeq 50\%$ smaller. We also note that rate coefficients for pure rotational deexcitation of H$_{2}$ by He were studied by~\citet{Zhou_2017} who conducted quantum scattering calculations on the BSP PES~\citep{Bakr2013}. From Fig.~5 in their paper, we can deduce that rate coefficients for the ${v=0,j=2\rightarrow v'=0,j'=0}$ transitions calculated in the range 25 to 150~K are slightly lower than the experimental datapoints, and are closer to the results of~\citet{Balakrishnan_1999a}.

{Now looking at transitions with $v > 0$, in} general, we obtain a good agreement with the results of~\citet{Flower_1998} and \citet{Balakrishnan_1999a} for transitions with $\Delta v = 0$, and $\Delta j = \pm 2, \pm 4$. As the difference in vibrational and rotational quantum numbers increases, we observe significant differences between our results and previous theoretical calculations. In particular, as shown in the bottom left (for deexcitation from the ${v=1,j=0}$ state) and bottom right (for deexcitation from the ${v=2,j=0}$ state) {of Fig.~\ref{fig5}}, we observe that rate coefficients for vibrational deexcitation are significantly (1-2 orders of magnitude) lower than the results of~\citet{Flower_1998} and \citet{Balakrishnan_1999a}. We attribute the discrepancies between the results to the quality of the PESs. We recall that the most recent PES was calculated using coupled-cluster method with single, double, and perturbative triple excitations [CCSD(T)], supplemented by full configuration interaction corrections. On the other hand, the MR PES was obtained using much smaller Gaussian basis sets and a lower level of electronic structure theory. Moreover, the BSP3 PES covered a significantly larger range of intramolecular distances ($\rhh\in [0.65, 3.75]$~$a_{0}$), which is crucial for the accuracy of quantum scattering calculations involving vibrationally excited H$_{2}$.

\section{Astrophysical applications}
\label{sec:discussion}

\begin{figure}[!ht]
    \centering
    \includegraphics[width=0.9\linewidth]{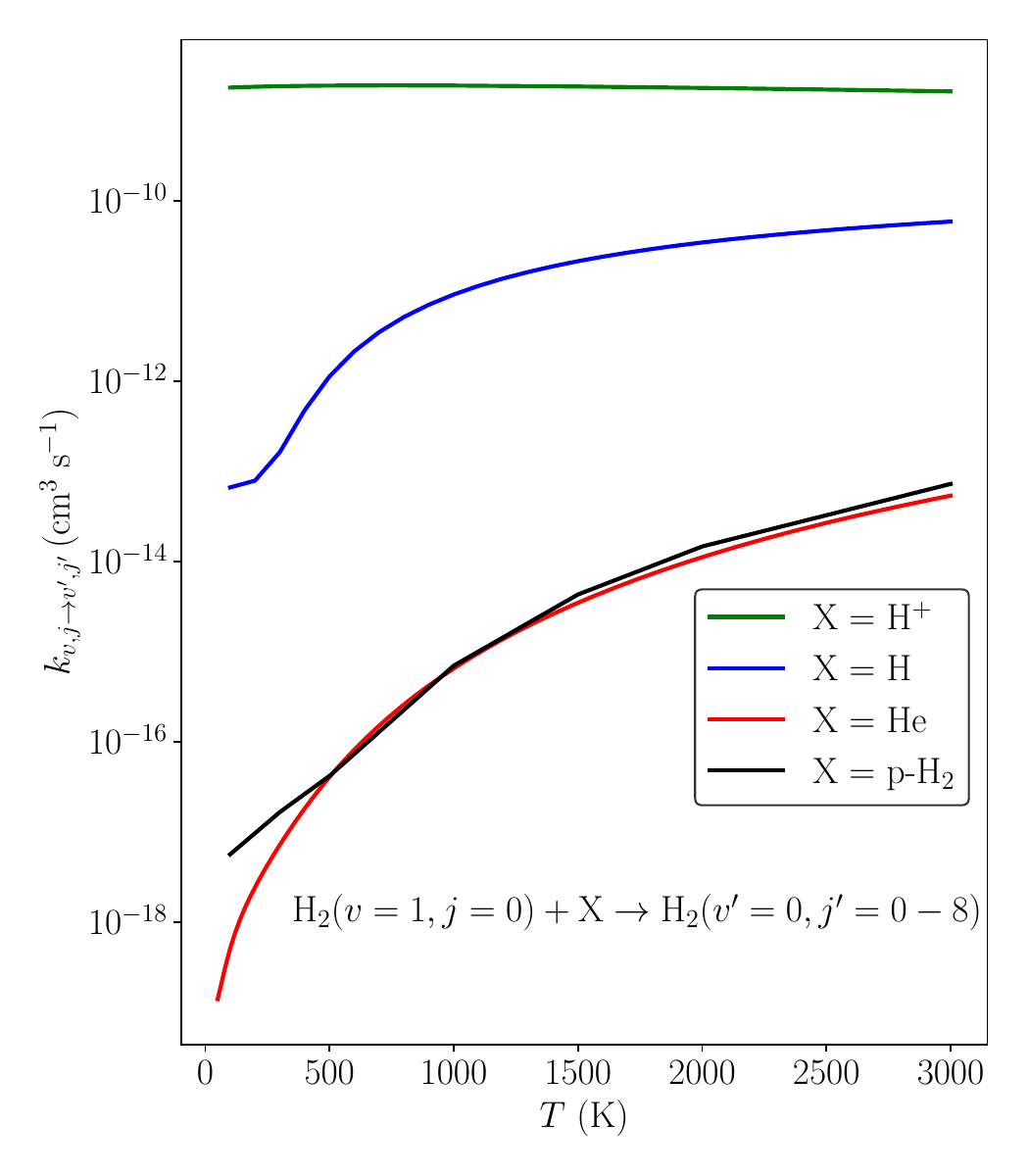}
    \caption{Comparison between the rate coefficients for rovibrational {$v=1, j=0 \rightarrow v'=0,\sum_0^8 j'$} de-excitation provided in this work and available literature data (see text).}
    \label{fig6}
\end{figure}

As {previously} mentioned, in the early Universe, as well as in interstellar clouds, the H$_2$ molecules can be excited by collisions with He, H, H$_2$ and H$^+$ which are the dominant heavy projectiles in such media.
It is then of interest to compare and discuss the efficiency of all projectiles for (de-)exciting H$_2$ by collisions.
Hence, we compare in Fig.~\ref{fig6} the new H$_2$--He collisional data for the rovibrational relaxation of H$_2(v=1,j=0)$ to H$_2(v'=0,\sum_0^8 j')$ to those for H$_2$--H \citep{Lique:15}, H$_2$--H$^+$ \citep{Gonzalez:21} and H$_2$--p-H$_2$ \citep{Flower_1998b} as a function of the temperature.

As one can see, H$^+$ is by far the most efficient projectile to \linebreak(de-)excite H$_2$, especially at low temperatures. Such dominance is not surprising and can be explained by the charge of H$^+$ leading to a strong interaction between H$_2$ and H$^+$ and hence to an efficient energy transfer during collisions. 

For the (de-)excitation induced by neutral projectiles, H is significantly dominating over He and H$_2$ because of its lighter mass and the reactive nature of the system (H$_2$ can be excited by H via inelastic and reactive processes, \cite{Lique:15}). Over all the temperature range explored in this work, He- and p-H$_2$-rate coefficients exhibit similar magnitude showing that the He:H$_2$ abundance ratio will not be a crucial parameter for determining the excitation conditions of H$_2$ in both the early universe and in interstellar clouds.

Then, as a second application and in order to test the impact of the new rate coefficients compared to those of F98 widely used in the astrophysical community, we perform radiative transfer calculations in order to determine the population of H$_2$ induced by He collisions. Non-local thermodynamic equilibrium (non-LTE) calculations were performed with the \texttt{RADEX} code \citep{vandertak_07}. Both collisional and radiative processes are taken into account. {Only the He projectiles have been taken into account.} We simulate the excitation conditions of H$_2$ induced by He for physical conditions corresponding to the early Universe \citep{Flower_2021} and typical warm molecular clouds where ro-vibrational lines of H$_2$ are detected \citep{Neufeld_2008}. Tables~\ref{tab1} and \ref{tab2} compare the H$_2$ fractional population, $p_{v,j}$, obtained using present and F98 collisional data for a temperature of 5000~K and a He density of 10$^3$~cm$^{-3}$ (typical physical conditions in the early Universe) and for a temperature of 1000~K and a He density of 10$^5$~cm$^{-3}$ (typical physical conditions in warm molecular clouds), respectively. 

\begin{table}[!ht]
\begin{center}
\caption{Comparison between H$_2$ fractional population, $p_{v,j}$, obtained using present and F98 collisional data.}
\label{tab1}
\small
\begin{tabular}{c|cc}
& \multicolumn{2}{c}{$T = 5000$~K, $n({\rm He})=10^3$~cm$^{-3}$} \\
Level & $p_{v,j}$ (This work) & $p_{v,j}$ (F98) \\
 \hline
$v=0, j=2$ & 2.542 $\times$ 10$^{-1}$ & 2.381 $\times$ 10$^{-1}$ \\
$v=0, j=4$ & 3.585 $\times$ 10$^{-1}$ & 3.405 $\times$ 10$^{-1}$ \\
$v=0, j=6$ & 2.648 $\times$ 10$^{-1}$ & 2.753 $\times$ 10$^{-1}$ \\
$v=1, j=0$ & 6.574 $\times$ 10$^{-6}$ & 2.470 $\times$ 10$^{-5}$ \\
$v=1, j=2$ & 3.646 $\times$ 10$^{-5}$ & 1.233 $\times$ 10$^{-4}$ \\ 
$v=1, j=4$ & 3.646 $\times$ 10$^{-5}$ & 1.233 $\times$ 10$^{-4}$ \\
$v=2, j=0$ & 9.924 $\times$ 10$^{-9}$ & 1.641 $\times$ 10$^{-7}$ \\
$v=2, j=2$ & 6.026 $\times$ 10$^{-8}$ & 8.365 $\times$ 10$^{-7}$ \\
$v=3, j=0$ & 4.233 $\times$ 10$^{-11}$ & 3.060 $\times$ 10$^{-9}$ \\
\hline \hline
\end{tabular}
\end{center}
\end{table}

\begin{table}[!ht]
\begin{center}
\caption{Comparison between H$_2$ $p_{v,j}$ obtained using present and F98 collisional data.}
\label{tab2}
\small
\begin{tabular}{c|cc}
& \multicolumn{2}{c}{$T = 1000$~K, $n({\rm He})=10^5$~cm$^{-3}$} \\
Level & $p_{v,j}$ (This work) & $p_{v,j}$ (F98) \\
 \hline
$v=0, j=2$ & 4.951 $\times$ 10$^{-1}$ & 4.942 $\times$ 10$^{-1}$ \\
$v=0, j=4$ & 2.753 $\times$ 10$^{-1}$ & 2.749 $\times$ 10$^{-1}$ \\
$v=0, j=6$ & 6.106 $\times$ 10$^{-2}$ & 6.191 $\times$ 10$^{-2}$ \\
$v=0, j=8$ & 3.612 $\times$ 10$^{-3}$ & 4.337 $\times$ 10$^{-3}$ \\
$v=1, j=0$ & 7.374 $\times$ 10$^{-8}$ & 2.692 $\times$ 10$^{-7}$ \\
$v=1, j=2$ & 2.425 $\times$ 10$^{-7}$ & 8.941 $\times$ 10$^{-7}$ \\
$v=1, j=4$ & 2.257 $\times$ 10$^{-7}$ & 7.768 $\times$ 10$^{-7}$ \\
$v=1, j=6$ & 1.370 $\times$ 10$^{-7}$ & 2.257 $\times$ 10$^{-7}$ \\
\hline \hline
\end{tabular}
\end{center}
\end{table}

As one can see the population of the different rotational levels in $v=0$ vibrational manifold are weakly impacted by the use of the new rate coefficients. A deviation of less than 5\% is typically observed for these levels. Such a weak impact can be explained by the good overall agreement between present and F98 pure rotational data. When {the vibrational quantum number increases}, substantial differences, larger than an order of magnitude, start to appear. These significant deviations reflect the large deviation between present and F98 rovibrational rate coefficients. Globally, the population of rovibrationally excited levels of H$_2$ induced by He collisional is weaker when using present data than when using F98 ones, reflecting new collisional data weaker in magnitude than the former ones.

The magnitude of the line intensities of rovibrational transitions detected by telescopes is directly proportional to the population of the energy levels. The significant deviations seen are likely to modify the observations analysis and the final determination of the abundance of H$_2$ in molecular clouds, even if He is not the dominant collider. The population of $v=1$ energy levels when using the new data are a factor 3-5 lower than when using the F98 data. We anticipate an increase of the H$_2$ abundance derived from the observations since modeling observational spectra with a weaker population will require to increase in the H$_2$ column density.

Such differences are also likely to modify the cooling function of H$_2$ that is strongly dependent on the H$_2$ populations  \citep{Flower_2021}. Nevertheless, we note that the largest deviations in the fractional populations are seen for weakly populated levels and we anticipate that the global impact on the cooling function will be moderate. 

\section{Conclusions}
\label{sec:conclusions}
We used the state-of-the-art PES~\citep{Thibault_2017} to perform quantum scattering calculations {for the} H$_{2}$-He system and to revise the state-to-state rate coefficients calculated by~\citet{Flower_1998}. The revised rates show consistency with previous studies for pure rotational transitions within the ground vibrational state as reported by~\citet{Flower_1998} and~\citet{Balakrishnan_1999a}, but significant discrepancies emerge for rovibrational transitions involving highly-excited rotational and vibrational states. We attribute {these discrepancies} to the superior accuracy of the PES and the broader range of intramolecular distances in H$_{2}$ covered by \textit{ab initio} points, which is crucial for accurate description of inelastic processes involving excited rovibrational states.

The new collisional data have been introduced in a radiative transfer code in order to simulate the excitation conditions of H$_2$ in the early Universe and in warm molecular clouds. We have found that the population of rotational levels in $v=0$ vibrational manifold is weakly impacted by the use of the new collisional data. On the opposite, the population of rotational levels in $v>0$ vibrational manifold is significantly reduced demonstrating the need of using the new set of data in astrophysical models.

\section*{Acknowledgements}

We acknowledge financial support from the European Research Council (Consolidator Grant COLLEXISM, Grant Agreement No. 811363), and the financial support of the University of Rennes via a grant project dedicated to international collaborations and via the CNRS IRN MCTDH grant. H. J. is supported by the Foundation for Polish Science (FNP) and by the National Science Centre in Poland through Project No. 2019/35/B/ST2/01118. P.W. is supported by the National Science Centre in Poland through Project No. 2022/46/E/ST2/00282. For the purpose of Open Access, the author has applied a CC-BY public copyright license to any Author Accepted Manuscript (AAM) version arising from this submission.

\bibliographystyle{aa}
\bibliography{bibliography.bib}
\end{document}